\documentclass[aps,pra,twocolumn]{revtex4}

\usepackage[usenames,dvipsnames]{color}

\usepackage[latin1]{inputenc}
\usepackage{graphicx}
\usepackage{amssymb}
\usepackage{subfigure}
\usepackage{stackrel}

\usepackage{amsmath}
\usepackage{amsfonts}
\usepackage{bm}
\usepackage{color}
\usepackage{verbatim} 

\newcommand{\ket}[1]{\ensuremath{\left|{#1}\right\rangle}}
\newcommand{\bra}[1]{\ensuremath{\left\langle{#1}\right |}}

\newcommand{\beq}{\begin{equation}}
\newcommand{\eeq}{\end{equation}}
\newcommand{\bse}{\begin{subequations}}
\newcommand{\ese}{\end{subequations}}
\newcommand{\bea}{\begin{eqnarray}}
\newcommand{\eea}{\end{eqnarray}}
\newcommand{\bit}{\begin{itemize}}
\newcommand{\eit}{\end{itemize}}
\newcommand{\bpmatrix}{\begin{pmatrix}}
\newcommand{\epmatrix}{\end{pmatrix}}

\newcommand{\be}{\begin{equation}}
\newcommand{\ee}{\end{equation}}
\newcommand{\ben}{\begin{eqnarray}}
\newcommand{\een}{\end{eqnarray}}

\begin{document}

\title{Multipartite concurrence for identical-fermion systems}

\author{A. P. Majtey$^{1,2,3}$, P. A. Bouvrie$^{4,5}$, A. Vald\'es-Hern\'andez$^6$, and A. R. Plastino$^{2,7}$}
\affiliation{$^1$Facultad de Matem\'atica, Astronom\'{\i}a y F\'{\i}sica, Universidad Nacional de C\'ordoba, Av. Medina Allende s/n, Ciudad Universitaria, X5000HUA C\'ordoba, Argentina}
\affiliation{$^2$Consejo de Investigaciones Cient\'{i}ficas y T\'ecnicas de la Rep\'ublica Argentina, Av. Rivadavia 1917, C1033AAJ, CABA, Argentina}
\affiliation{$^3$Instituto de F\'{\i}sica, Universidade Federal do Rio de Janeiro, Caixa
Postal 68528, Rio de Janeiro, RJ 21941-972, Brazil}
\affiliation{$^4$Centro Brasileiro de Pesquisas F\'isicas, Rua Dr. Xavier Sigaud 150, Rio de Janeiro, RJ 22290-180, Brazil}
\affiliation{$^5$Instituto Carlos I de F\'isica Te\'orica y Computacional, Universidad de Granada, E-18071 Granada, Spain}
\affiliation{$^6$Instituto de F\'{\i}sica, Universidad Nacional Aut\'{o}noma de M\'{e}xico, \\
Aparatado Postal 20-364, M\'{e}xico D.F., Mexico}
\affiliation{$^7$CeBio  y Secretar\'{i}a de Investigaciones, Universidad Nacional del Noroeste de la Prov. de Buenos Aires, UNNOBA-Conicet, Roque Saenz-Pe\~{n}a 456, Junin, Argentina}

\email[]{amajtey@famaf.unc.edu.ar}

\begin{abstract}

We study the problem of detecting multipartite entanglement among indistinguishable  fermionic particles. A multipartite concurrence for pure states of $N$ identical fermions, each one having a $d$-dimensional single-particle Hilbert space, is introduced. Such entanglement measure, in particular, is optimized for maximally entangled states of three identical fermions that play a role analogous to the usual (qubit) Greenberger-Horne-Zeilinger-state. In addition, it is shown that the fermionic multipartite concurrence can be expressed as the mean value of an observable, provided two copies of the composite state are available. 

\end{abstract}

\maketitle

\section{Introduction}

Entanglement is both a central key for understanding quantum phenomena and a useful resource for the implementation of quantum information tasks \cite{BZ06,AFOV08}. Identical particles, on the other hand, are essential for understanding the properties of many-particle quantum systems \cite{MG64}. For systems of identical particles, fermions or bosons, however, even the very notion of entanglement is controversial \cite{BFT14}. In the fermionic case, there exists some extended consensus that a pure fermion state is separable if it is a single antisymmetric product state given by a single Slater determinant \cite{ESBL02, GMW02, GM04, SLM01, SCKLL01, PMD09,RSV15}. In this paper we adopt such a point of view and consider entanglement in systems of identical fermions, meaning entanglement between particles and not entanglement between modes \cite{GR15}. 

Several efforts have been devoted to the study of the entanglement features in systems of $N$ identical fermions in the last few years \cite{NV07,BMPSD2012,LV08, BPCP08, LNP05, ZP10, GM05, OSTS08, OK13, BFM14}. Various bipartite entanglement measures for pure $N$-fermion states have been discussed, yet these measures are, in general (for $N>2$), difficult to implement \cite{ESBL02}. An interesting (bi)separability criterion for systems of $N$ identical fermions was formulated in \cite{PMD09}. 

When studying multipartite entanglement, it is convenient to consider all possible bipartitions of the complete system. Then, the available entanglement measures for bipartite systems become applicable, and adequate generalizations can account for real multipartite correlations \cite{CMB04, MKB05}.

In the present contribution we propose a multipartite concurrence measure for fermionic systems in a pure state, and analyze its main properties as a suitable measure of multipartite entanglement. An appropriate criterion of bipartite entanglement, valid for arbitrary bipartitions of the system, is also obtained, as well as the fermionic analog of the standard Greenberger-Horne-Zeilinger (GHZ) states, characterized by possessing maximal multipartite entanglement, and by the fact that tracing over one of the subsystems destroys any entanglement present among the constituents. Furthermore, we explicitly show how the concurrence measure can be written in terms of the mean value of an observable given that a twofold copy of the state in question is available.  

This work is structured as follows. Section II contains the preliminaries for the subsequent construction of the multipartite concurrence measure for pure states of $N$ indistinguishable fermions. First, we briefly outline the main features of the concurrence for distinguishable-party systems. Then we introduce the definition of separability in fermionic systems and present a concurrence measure for a pure state with $N=2$.  In Sec. III we present a general separability criterion for arbitrary bipartitions $(M: N-M; 1\leq M\leq N-1)$, and introduce suitable bipartite and multipartite concurrence measures for arbitrary $N$. In Sec. IV we present two observables whose mean value (provided two copies of the fermionic system are available) coincides with the fermionic multipartite concurrence. Finally, some conclusions are drawn in Sec. V.

\section{Preliminaries}

\subsection{Concurrence in distinguishable-party system}

The concurrence was first introduced in \cite{HW97} as a measure of the entanglement between two qubits, having a one-to-one correspondence with entanglement of formation \cite{W98}. The measure was then generalized to $(d_A\times d_B)$-dimensional bipartite pure states $\psi_{AB}$ according to \cite{PRA64} 
\begin{equation}
\label{bipConcurrence}
C_{AB}=C(\psi_{AB})=\sqrt{2(1-\textrm{Tr}\rho_A^2)}=\sqrt{2(1-\textrm{Tr}\rho_B^2)}
\end{equation}
(we wrote $\langle\psi|\psi\rangle=1$). Now, when considering $N$-partite systems, multipartite correlations between subsystems may appear. For an $N$-partite pure state $\psi_{N}$ a suitable generalization of Eq. (\ref{bipConcurrence}) is the so-called multipartite concurrence \cite{CMB04}

\begin{equation}
\label{MultipartiteConcurrence}
C_N=C(\psi_N)=2^{1-(N/2)}\sqrt{(2^{N}-2)-\textrm{Tr}\sum_i\rho_i^2},
\end{equation}
where the index $i$ labels all $(2^N-2)$ subsets of the $N$-particle system, and $\rho_i$ are the reduced density matrices of all one- to $(N-1)$-partite subsystems \cite{MKB05}. For $\ket{\psi_{N}}=\ket{\psi_{N-1}}\otimes \ket{\phi}$ the multipartite concurrence becomes $C_N(\psi_{N})=C_{N-1}(\psi_{N-1})$, hence $C_N$ vanishes for fully separable states $\ket{\psi_{N}}=\otimes^{N}_{i=1}\ket{\phi_{i}}$. Moreover $C_N$ reaches its maximum value for GHZ states $\ket{\psi_{N}}=\sum_{i}\ket{i...i}/\sqrt{2}$.

The concurrence $C_N$ can be expressed as the following expectation value with respect to two copies of the system \cite{MKB05}:
\begin{equation}\label{ConOpA}
C_N=\sqrt{\langle \psi_N|\otimes\langle\psi_N| A|\psi_N\rangle\otimes|\psi_N\rangle}
\end{equation}
where
\begin{equation}
\label{ObservableMKB}
A=4\sum_{\{s_{j_i}=\pm\}^+}P_{s_{1_i}}^1\otimes...\otimes P_{s_{N_i}}^N,
\end{equation}
and 
\begin{equation}
\label{ProjectorsMinter}
 P_{\pm}^{i}=\frac{1}{4}\sum_{\alpha_i,\alpha_i'}(\ket{\alpha_i} \ket{\alpha_i'} \pm \ket{\alpha_i'} |\alpha_i\rangle)(\langle \alpha_i| \bra{\alpha_i'} \pm \langle \alpha_i'| \bra{\alpha_i})
\end{equation}
with $1\leq i\leq N$. Here $P_+^i$ and $P_-^i$  are the projectors onto the symmetric $(+)$ and antisymmetric $(-)$ subspaces of the Hilbert space $\mathcal{H}_j\otimes \mathcal{H}_j$ that describes the two copies of the $j$-th subsystem. The sum in Eq. (\ref{ObservableMKB}) is restricted to the set $\{s_{j_i}=\pm\}^+$ composed of all possible ways of sorting the symbols $+$  and $-$, with an even number of $-$ symbols, and excluding the completely symmetric case with no $-$ symbols at all. 

In \cite{AM2006} it was argued that the observable $A$ can be replaced by the single factorizable observable $\tilde A$ given by
\bea
\label{ObservableAM}
\tilde A=4 ( \mathbb I - P_{+}^1\otimes...\otimes P_{+}^N).
\eea
In this way $C_N$ can be constructed in a much more efficient way, and can be experimentally determined measuring only one single probability \cite{AM2006, WRDMB06}.

\subsection{Definition of separability in systems of identical fermions}

Let us now consider a system composed of $N$ indistinguishable fermions, each one having a $d$-dimensional single-particle orthonormal basis $\mathcal{B}=\{\ket{1},\ldots,\ket{d}\}$. We introduce the fermionic creation operators $f_i^{\dagger}$ ($1\leq i\leq d$) acting on the fermionic vaccuum $|0\rangle$ as those that produce the totally antisymmetric combination
\bea
\label{AntisymmetricState}
\ket{\psi^{sl}_i} &=& \hat f_{i_1}^\dagger \cdots \hat f_{i_N}^\dagger \ket{0}, 
\eea
where $\ket{\psi^{sl}_i}$ is a Slater determinant 
\bea
\label{slater}
\ket{\psi^{sl}_i} &=& \frac{1}{\sqrt{N!}} \sum_{P\{i\}} \varepsilon^{i_1\ldots i_N} \ket{{i_1}{i_2}\ldots{i_N}}. 
\eea
Here $P\{i\}$ are the $N!$ permutations of the set $\{i_1,\ldots,i_N\}$, and $\varepsilon^{i_1\ldots i_N}$ stands for the $N$-dimensional totally antisymmetric unitary tensor. Notice that in order to construct an antisymmetric $N$-fermion state we must have $N\leq d$.

Although, clearly  (\ref{slater}) is a nonfactorizable state, in this paper we stick to the extended consensus that in systems of identical fermions the minimum quantum correlations between the particles that are required by the indistinguishability and the antisymmetry of the fermionic state do not contribute to the state's entanglement \cite{AFOV08, ESBL02, GMW02, GM04, SCKLL01, PMD09, NV07,BMPSD2012,LV08, BPCP08, LNP05, ZP10, GM05, OSTS08}. Therefore, in what follows a composite system of $N$ identical fermions is regarded as separable (i.e., nonentangled) if and only
if its density matrix can be expanded as \cite{GMW02}
\begin{equation}
\rho^{sep}=\sum_{k}p_{k}\ket{\psi^{sl}_k}\bra{\psi^{sl}_k}
\label{rhosep},
\end{equation} 
with $\ket{\psi^{sl}_k}$ being a Slater determinant (said to have Slater rank 1), and $\sum_{k}p_{k}=1.$ That is, a pure separable state of $N$ identical fermions is simply a single Slater determinant, whereas mixed separable states are those that can be expressed as a statistical mixture of pure states of Slater rank 1.

\subsection{Bipartite concurrence in systems of two identical fermions}
Equation (\ref{rhosep}) already indicates that quantification of entanglement in identical-fermion systems exhibits some differences from the corresponding concept as applied to systems consisting of distinguishable subsystems. The lowest-dimensional system allowing a Slater rank larger than 1, hence allowing entanglement, has $N=2$ and $d=4$, thus resulting in a six-dimensional two-particle Hilbert space. For this particular system, a fermionic analog of the two-qubit concurrence exists, which measures the fermion-fermion entanglement \cite{SLM01,ESBL02}. For an arbitrary pure state of the two fermions,  
\begin{equation}\label{psiff}
 |\psi \rangle=\sum_{i,j=1...4} w_{ij} ~ \hat f_i^{\dagger} \hat f_j^{\dagger}|0\rangle,
\end{equation}
with $w_{ij}$ being the elements of an antisymmetric matrix $w$ that fulfills the normalization condition $\textrm{Tr}(w w^{\dagger})=1/2$, the fermionic concurrence reads 

\begin{equation}
C_{ff}(\psi)=8|w_{12}w_{34}-w_{13}w_{42}+w_{14}w_{23}|,
\end{equation}
which in turn can be expressed as [see Eq. (\ref{bipConcurrence})]
\begin{equation}\label{CF}
C_{ff}(\psi)=\sqrt{2(1-2\textrm{Tr}\rho_f^2)},
\end{equation}
where $\rho_f$ is the single-fermion reduced density matrix. 

Recently \cite{VHMP2015}, a tripartite system was considered that involved a pair of indistinguishable fermions and a third party $A$ (arbitrary except for that no third identical fermion is contained in it). For a pure state $\ket{\phi}$ of such a tripartite system, a measure of entanglement defined in terms of the purity of the fermionic reduced density matrix has been proposed to quantify the bipartite entanglement between one of the fermions and the rest of the system (second fermion plus $A$) \cite{VHMP2015}. Now, for states of the form $\ket{\phi}=\ket{\psi}_{ff}\ket{\eta}_{A}$, clearly the entanglement between one fermion and the rest reduces to the entanglement between the fermions (which are in a pure state). This means that the aforementioned measure can be considered a suitable concurrence for any two-fermion pure state, that is
\begin{equation}\label{CAAA}
C_{ff}(\psi)= \sqrt{\frac{2d}{d-2}\left(\frac{1}{2}-\textrm{Tr}\rho^{2}_{f}\right)},
\end{equation}
in total analogy with the concurrence involving distinguishable subsystems, Eq. (\ref{bipConcurrence}). Notice that for $d=4$, Eq. (\ref{CAAA}) reduces to Eq. (\ref{CF}), as expected. The factor ${2d}/({d-2})$ normalizes the concurrence, so that $C_{ff}=1$ corresponds to a maximally entangled state.

In the following section we will be interested in a generalization of Eq. (\ref{CAAA}) to pure states of $N$ identical fermions. This will be of use in the generalization of the multipartite concurrence (\ref{MultipartiteConcurrence}) to fermionic systems.

\section{Multipartite fermion entanglement}

\subsection{General entanglement criterion for pure $N$-fermion states}
\label{ReviewEntNFermions}

A convenient bipartite entanglement criterion for pure states of systems of $N$ identical fermions was introduced in \cite{PMD09}. It can be formulated in terms of the purity $\textrm{Tr}\rho_1^2$ of a single-fermion reduced density matrix, and reads
\begin{equation}
\label{entcritEPL}
\begin{cases} \textrm{Tr}\rho_1^2=\frac{1}{N}&\text{nonentangled,}\\ \frac{1}{d}\leq\textrm{Tr}\rho_1^2<\frac{1}{N}& \text{entangled.} \end{cases}
\end{equation}
In previous sections the reduced density matrix of a single fermion was denoted as $\rho_f$. From now on, and for clarity purposes, we will denote with $\rho_M$ the reduced density matrix of a subsystem containing $M$ fermions (with $1\leq M \leq N-1)$, i.e., $\rho_M=\textrm{Tr}_{(M+1,\ldots,N)}\rho$.  

A generalization of the entanglement criterion \eqref{entcritEPL} that holds for arbitrary bipartitions $M:N-M$ of the complete system is necessary in order to construct a multipartite concurrence measure, analogous to Eq. (\ref{MultipartiteConcurrence}), valid for identical fermions. 
For an $N$-fermion pure state $\rho=\ket{\psi}\bra{\psi}$, such a generalization can be formulated in terms of the purity of $\rho_M$, which fulfills

\bea
\label{TrRhoM2Inequality}
\text{Tr}\rho_{M}^2 \leq \binom{N}{M}^{-1}.
\eea
We provide a complete proof of inequality \eqref{TrRhoM2Inequality} in Appendix \ref{TraceRhoMProof}. We also show that the equal sign holds if and only if $\ket{\psi}$ has Slater rank 1, so the state is separable. This allows us to generalize (\ref{entcritEPL}) as follows

\begin{equation}
\label{entcritBouvrie}
\begin{cases} \textrm{Tr}\rho_M^2=\binom{N}{M}^{-1} &\text{nonentangled,}\\ \frac{1}{d_M}\leq\textrm{Tr}\rho_M^2<\binom{N}{M}^{-1} & \text{entangled,} \end{cases}
\end{equation}
where $d_M=\binom{d}{\textrm{min}\{M,N-M\}}$. 

\subsection{Multipartite concurrence}

The entanglement criterion (\ref{entcritBouvrie}) allows us to formulate an appropriate fermionic multipartite concurrence $C_{N_f}$ by demanding that it vanishes whenever all the reduced density matrices $\rho_{M}$, corresponding to all possible subsystems, are minimally mixed.

From Eq. \eqref{TrRhoM2Inequality} we find that 
\bea
\sum^{N-1}_{M=1}\binom{N}{M}\text{Tr}\rho_M^2 \leq \sum^{N-1}_{M=1}1=(N-1),
\eea
where the equal sign holds only for separable states. Therefore the quantity
\bea
\label{ConcuNonNormalized}
C_{N_f} (\psi) =\sqrt{\alpha_{N}\big{[}(N-1)-\sum^{N-1}_{M=1}\binom{N}{M}\text{Tr}\rho_M^2\big{]}},
\eea 
with $\alpha_{N}\geq 0$, can be considered a suitable multipartite concurrence, analogous to Eq. (\ref{MultipartiteConcurrence}), for the $N$-fermion system. The factor $\alpha_{N}$ is fixed depending on the maximum value allowed for $C_{N_f}$. By setting the maximal entanglement equal to unity ($C_{N_f}\leq 1$), we are led to 
\begin{equation}
\alpha_{N}=\frac{1}{(N-1)-\sum^{N-1}_{M=1}\binom{N}{M}\frac{1}{d_M}}.
\end{equation}
With Eq. (\ref{ConcuNonNormalized}) at hand we are in a position to investigate whether the maximum value $C_{N_f}=1$ is actually achieved for some multi-fermionic states. A more detailed investigation of the emergence of maximally multipartite-entangled pure states in systems of $N$ identical fermions is left for future analysis. Here it suffices to consider three fermions with a single-particle Hilbert space of dimension 6 in the following state:
\begin{equation}\label{3fermion}
|\psi \rangle_{fGHZ}=\frac{1}{\sqrt{2}}(\hat f_{1}^{\dagger} \hat f_{2}^{\dagger} \hat f_{3}^{\dagger}|0\rangle+\hat f_{4}^{\dagger} \hat f_{5}^{\dagger} \hat f_{6}^{\dagger}|0\rangle).
\end{equation}
Direct calculation shows that for such a state $C_{N_f}=1$. Moreover, the reduced two-fermion density matrices correspond to separable states [of the form (\ref{rhosep})], so that the tripartite state is maximally entangled whereas tracing over any one of the subsystems destroys any entanglement present. The fact that this last property is characteristic of the three-qubit GHZ states explains the subindex in $|\psi \rangle_{fGHZ}$, stressing that the latter is the fermionic version of the usual GHZ states.  
\section{Concurrence as the mean value of an observable}
In this section we show that the above multipartite concurrence \eqref{ConcuNonNormalized} can be expressed as the mean value of an observable, provided two (distinguishable) copies of the composite state are available. 

\subsection{Observable related to the linear entropy}

Let $\ket{\psi}_{AB}$ be a bipartite pure state. In this section $A$ and $B$ may have an arbitrary number of subsystems (distinguishable or not) of arbitrary dimensions. 
The density matrix of the composite system is $\rho=\ket{\psi}\bra{\psi}$, and the reduced density matrix of $A$ reads
\bea
\label{rhoA}
\rho_A &=& \textrm{Tr}_{B}\ket{\psi}\bra{\psi}=\sum_{\beta}\ket{\phi_\beta}\bra{\phi_\beta},
\eea
where $\ket{\phi_\beta}=\langle \beta| \psi\rangle$, with $\{\ket{\beta}\}$ being an orthonormal basis of $\mathcal{H}_B$.
Equation \eqref{rhoA} gives
\beq
\rho^{2}_A = \sum_{\beta \beta^{\prime}}\langle \phi_\beta|\phi_{\beta^{\prime}}\rangle \ket{\phi_\beta}\bra{\phi_{\beta^{\prime}}},
\eeq
so
\begin{eqnarray}\label{tr}
\textrm{Tr}\rho^{2}_{A} &=& \sum_{\beta \beta^{\prime}}\langle \phi_\beta|\phi_{\beta^{\prime}}\rangle 
\langle \phi_{\beta^{\prime}}|\phi_{\beta}\rangle \nonumber \\
&=&\sum_{\beta \beta^{\prime}}\langle \psi|\beta \rangle\langle \beta^{\prime}|\psi \rangle\langle \psi|\beta^{\prime} \rangle\langle \beta|\psi \rangle \nonumber \\
&=&{\bra{\psi}_1} \otimes \bra{\psi}_2 \{\sum_{\beta \beta^{\prime}}\ket{\beta_1}\bra{\beta_1^{\prime}}\ket{\beta_2^{\prime}}\bra{\beta_2}\} \ket{\psi}_1 \otimes{\ket{\psi}_2} \nonumber \\
&=&{\bra{\psi}_1} \otimes \bra{\psi}_2 O_B \ket{\psi}_1 \otimes{\ket{\psi}_2}, 
\end{eqnarray}
where we introduced the subindices 1 and 2 to refer to the copies of the system. Notice that the operator
\begin{eqnarray}
O_B&=&\sum_{\beta \beta^{\prime}}\ket{\beta_1}\bra{\beta_1^{\prime}}\ket{\beta_2^{\prime}}\bra{\beta_2}
\end{eqnarray}
acts on only the two copies of subsystem $B$.

Let us now consider the projector operators in \eqref{ProjectorsMinter}
\begin{equation}
P_{\pm}^{(B)}= \sum_{\beta \beta^{\prime}} \frac{1}{4}(\ket{\beta}_1\ket{\beta^{\prime}}_2\pm \ket{\beta^{\prime}}_1\ket{\beta}_2)(\bra{\beta}_1\bra{\beta^{\prime}}_2\pm \bra{\beta^{\prime}}_1\bra{\beta}_2),
\end{equation}
which by direct calculation gives

\begin{eqnarray}
P_{\pm}^{(B)}
=\frac{\mathbb I\pm O_B}{2}.
\end{eqnarray}
Substituting into Eq. (\ref{tr}) we get (omitting unnecessary subindices)
\begin{equation}\label{Trfinal}
\textrm{Tr}\rho^{2}_{A}={\bra{\psi}} \otimes \bra{\psi}(\pm 2 P_{\pm}^{(B)} \mp \mathbb I) \ket{\psi} \otimes{\ket{\psi}}.
\end{equation}

Now, the linear entropy is $S=1-\textrm{Tr}\rho^{2}_A$, and because of the above results, we can conclude that any linear function of $S$ can be expressed as the mean value of an observable provided two copies of the bipartite system are available. As stated above, this holds for any pure state of arbitrary dimensions (qubits, qudits, distinguishable, indistinguishable, fermions, bosons, etc).  In particular, the usual concurrence (\ref{bipConcurrence}), or the fermionic concurrence (\ref{CAAA}), admits an expression in terms of an observable.  

In the expression for the multipartite concurrence (\ref{ConcuNonNormalized}) all reduced density matrices, or equivalently, all bipartitions $M:N-M$, were considered. Therefore it is convenient to rewrite Eq. (\ref{Trfinal}) as
\begin{equation}\label{Trfinalb}
\textrm{Tr}\rho^{2}_{M}={\bra{\psi}} \otimes \bra{\psi}O^{(N-M)} \ket{\psi} \otimes{\ket{\psi}},
\end{equation}
where the operator $O^{(N-M)}$ acting on the two copies of the reduced $(N-M)$-particle system is
\begin{equation}\label{ONM}
O^{(N-M)}=\pm 2 P_{\pm}^{(N-M)} \mp \mathbb I.
\end{equation}
Direct inspection of Eqs. (\ref{ConcuNonNormalized}) and (\ref{Trfinalb}) leads to
\begin{equation}\label{concuA}
C_{N_f}=\sqrt{\langle \psi|\otimes\langle\psi| A_f |\psi \rangle\otimes|\psi\rangle},
\end{equation}
with
\bea
\label{Afermi}
A_f=\alpha_{N}\big{[}(N-1)\mathbb I-\sum^{N-1}_{M=1}\binom{N}{M}O^{(N-M)}\big{]}.
\eea 
\subsection{Observable related to the (usual) multipartite-concurrence observable}

The observable (\ref{Afermi}) is clearly one of (in principle) infinitely many observables $A_f$ that comply with Eq. (\ref{concuA}).  A second observable will now be derived, based on the observable found in \cite{MKB05} for the usual multipartite concurrence. 
From Eqs. (\ref{MultipartiteConcurrence}) and (\ref{ConOpA}) we have
\begin{equation}
\label{Obs2}
\langle \psi_N|\otimes\langle\psi_N| A|\psi_N\rangle\otimes|\psi_N\rangle=2^{2-N}[(2^{N}-2)-\textrm{Tr}\sum_i\rho_i^2],
\end{equation}
where, as stated in connection with Eq. (\ref{MultipartiteConcurrence}), the index $i$ labels all the $(2^N-2)$ subsets of the $N$-partite system. When dealing with indistinguishable fermions systems, many terms in the sum $\sum_i\rho_i^2$ are identical. Specifically, there are $\binom{N}{M}$ subsystems characterized by the same $\rho_M$. In the fermionic case, Eq. (\ref{Obs2}) is thus rewritten as 
\bea
\label{Obs2b}
\langle \psi_N|\otimes\langle\psi_N| A|\psi_N\rangle\otimes|\psi_N\rangle \nonumber \\ = 2^{2-N} (2^{N}-2) - 2^{2-N} \sum_{M=1}^{N-1}  \binom{N}{M} \textrm{Tr} \rho_M^2.
\eea
Comparison with Eq. (\ref{ConcuNonNormalized}) gives
\bea
C_{N_f}=\sqrt{\langle \psi_N| \otimes \langle \psi_N| A^{\prime}_f |\psi_N \rangle \otimes |\psi_N \rangle},
\eea
with
\bea
\label{FermionObserv}
A^{\prime}_f = \alpha_N  (1+N-2^N+2^{N-2} A),
\eea
and $A$ given by Eq. (\ref{ObservableMKB}). An immediate difference between the observable $A_f$ and $A^{\prime}_f$ is that the former involves $N-1$ operators, whereas the latter involves a single factorizable observable.

Experimentally, one usually faces a difference between the two copies of the state 
whose concurrence one wants to determine, due to limited precision in the
states' preparation procedure followed in the laboratory. 
In general, the two prepared copies will not match exactly. 
A similar difficulty arises, of course, when measuring
the entanglement of bipartite systems with distinguishable
subsystems \cite{WSRDMB07}. In order to discuss how sensitive 
the estimation of the concurrence [via equations such as (\ref{ConcuNonNormalized})] is  
to small deviations from the ideal preparation of identical copies, let us assume that we are 
dealing with two different (normalized) states $|\psi_N\rangle$ and $|\psi_N^{\prime}\rangle$, with 
$|\langle\psi_N|\psi_N^{\prime}\rangle|\lesssim1$. The state $|\psi_N^{\prime}\rangle$ 
can be expressed as

\be \label{fipri}
|\psi_N^{\prime}\rangle=\sqrt{1-\varepsilon^2}|\psi_N\rangle+\varepsilon|\delta\psi_N\rangle,
\ee
with $\langle\delta\psi_N|\delta\psi_N\rangle= 1$ and  
 $\langle\psi_N|\delta\psi_N\rangle=0$. As discussed in \cite{WSRDMB07}, the sensitivity of the measurement process to the mismatch of both copies of the state can be estimated by comparing $C_{\textrm{exp}}=\sqrt{\langle\psi_N|\otimes\langle\psi_N^{\prime}|A|\psi_N\rangle\otimes|\psi_N^{\prime}\rangle}$ with the mean value $C_{\rm mean}=\frac{1}{2}[C(\psi_N)+C(\psi_N^{\prime})]$. 
 Inserting the expression in the right hand side of 
 (\ref{fipri}) into the expressions defining $C_{\rm exp}$
 and $C_{\rm mean}$, and expanding in powers of the small parameter
$\varepsilon$, it can be readily verified that these two
 quantities coincide to first order in $\varepsilon$.  
 The difference between these quantities is of order $\varepsilon^2$.
 Consequently, to first order in $\varepsilon$ the quantity that
 one is actually measuring, $C_{\rm exp}$, is a meaningful
 entanglement measure: it represents the average between the concurrences of the two copies. Second order errors depend on the
 specific forms of the states $|\psi_N\rangle$ and $|\psi_N^{\prime}\rangle$ and, consequently, 
 on the specific experimental process through which these states are prepared. 
 These errors can be analyzed only in a case-by-case way. The effects of errors in the preparation process leading 
 to the production of mixed instead of pure states will also affect the concurrence measurement. 
 To study the impact of this kind of error, we would need first to extend our present results  
 on fermionic multipartite concurrence to mixed states of $N$ identical fermions. We plan to address this issue in a future work.

\vspace{0.5cm}
\section{Conclusions}
Summarizing, we have introduced a multipartite concurrence for arbitrary-dimensional $N$-fermion pure states. This goal has been achieved by generalizing a bipartite separability criterion to arbitrary bipartitions $M:N-M$. In addition, the generalization also provided a bipartite measure of entanglement for fermionic pure states when any bipartition is considered. In the case $N=3$ we identified maximally entangled fermionic states which become separable after tracing over one of the constituents subsystems, in total analogy with the standard GHZ state. Finally,  we have shown how the proposed concurrence can be written in terms of the mean value of two different observables,  assuming that two copies of the fermionic state are available.
 
\begin{acknowledgements}
A.P.M, and P.A.B. acknowledge the Brazilian agencies MEC,MCTI,CAPES,CNPq,FAPs for the financial support through the \textit{BJT Ci\^encia sem Fronteiras} Program. P.A.B. acknowledges support from the Spanish Project Grants No. FIS2014-59311-P co-financed by FEDER funds. A.V.H. gratefully acknowledges financial support from DGAPA, UNAM through project PAPIIT IA101816. A.R.P. acknowledges financial supports from grant 401512/2014-2 of CNPq.\end{acknowledgements}

\begin{appendix}

\begin{widetext}

\section{Proof of inequality \eqref{TrRhoM2Inequality}}
\label{TraceRhoMProof}

Given a single-particle orthonormal basis $\{|i\rangle, i=1,  \ldots, d\}$, an arbitrary pure $N$ fermion state can be written as
\be
\label{GeneralFermionSate}
|\psi\rangle=\sum_{i_1\ldots i_N} \omega_{i_1\ldots i_N}f^{\dagger}_{i_1}\ldots f^{\dagger}_{i_N}|0\rangle,
\ee
or equivalentely,
\be
|\psi\rangle=\frac{1}{\sqrt{N!}}\sum_{i_1\ldots i_N} \omega_{i_1\ldots i_N}\sum_{P\{i\}}\varepsilon^{i_1\ldots i_N}|i_1\ldots i_N\rangle,
\ee
where the (in general) complex coefficients $\omega_{i_1\ldots i_N}$ are antisymmetric in all indices and comply with the normalization condition

\be
\sum_{i_1\ldots i_N} |\omega_{i_1\ldots i_N}|^2=\frac{1}{N!}.
\ee

The $M$-fermion reduced density matrix, $\rho_M=\textrm{Tr}_{M+1\ldots N}|\psi\rangle\langle\psi|$, reads
\ben
\rho_M&=&\sum_{j_{M+1}\ldots j_N}\langle j_{M+1}\ldots j_N|\psi\rangle\langle\psi|j_{M+1}\ldots j_N\rangle\nonumber\\
&=&\frac{1}{N!}\sum_{j_{M+1}\ldots j_N}\sum_{i_1\ldots i_N}\sum_{l_1\ldots l_N} \omega_{i_1\ldots i_N} \omega^*_{l_1\ldots l_N}\sum_{P\{i\}}\sum_{P\{l\}}\varepsilon^{i_1\ldots i_N}\varepsilon^{l_1\ldots l_N}\langle l_1\ldots l_N|j_{M+1}\ldots j_N\rangle\langle j_{M+1}\ldots j_N|i_1\ldots i_N\rangle\nonumber\\
&=&\frac{1}{N!}\sum_{i_1\ldots i_N}\sum_{l_1\ldots l_N} \omega_{i_1\ldots i_N} \omega^*_{l_1\ldots l_N}\sum_{P\{i\}}\sum_{P\{l\}}\varepsilon^{i_1\ldots i_N}\varepsilon^{l_1\ldots l_N}\langle l_{M+1}\ldots l_N|i_{M+1}\ldots i_N\rangle|i_1\ldots i_M\rangle\langle l_1\ldots l_M|.
\een

Let $G_{k_1\ldots k_M}$ denote the diagonal elements of $\rho_M$. Then,

\ben
\label{diagonalrhom}
G_{k_1\ldots k_M}&=&\langle k_1\ldots k_M| \rho_M | k_1\ldots k_M\rangle\nonumber\\
\nonumber\\&=&\frac{1}{N!}\sum_{i_1\ldots i_N}\sum_{l_1\ldots l_N} \omega_{i_1\ldots i_N} \omega^*_{l_1\ldots l_N}\sum_{P\{i\}}\sum_{P\{l\}}\varepsilon^{i_1\ldots i_N}\varepsilon^{l_1\ldots l_N}\langle l_{M+1}\ldots l_N|i_{M+1}\ldots i_N\rangle\nonumber\\&\times&\langle k_1\ldots k_M|i_1\ldots i_M\rangle\langle l_1\ldots l_M|k_1\ldots l_M\rangle.
\een

The last line of Eq. (\ref{diagonalrhom}) sets $l_1\ldots l_M = i_1\ldots i_M$, then $P\{l\}$ contributes with $(N-M)!$ terms, and we finally obtain
\ben
G_{k_1\ldots k_M}=\frac{1}{N!}\sum_{i_1\ldots i_N} |\omega_{i_1\ldots i_N}|^2(N-M)!N!|\langle k_1\ldots k_M|i_1\ldots i_M\rangle|^2,
\een

\ben
\label{A7}
G_{k_1\ldots k_M}=\begin{cases} (N-M)!\sum_{i_1\ldots i_N} |\omega_{i_1\ldots i_N}|^2, &\text{if}\;\; k_1\ldots k_M\in (i_1\ldots i_N),\\0, &\text{otherwise.}\end{cases}
\een

Taking into account the symmetric character of $G_{k_1\ldots k_M}$ under permutation of its indices, we can define

\ben
g_{k_1\ldots k_M}^{(i_1\ldots i_N)}\begin{cases} \frac{M!(N-M)!}{N!}, &\text{if}\;\; k_1<\ldots <k_M\in (i_1\ldots i_N),\\0, &\text{otherwise,}\end{cases}
\een
so Eq. (\ref{A7}) is finally rewritte nas
\ben
\label{A9}
G_{k_1\ldots k_M}=N!\sum_{i_1\ldots i_N} |\omega_{i_1\ldots i_N}|^2g_{k_1\ldots k_M}^{(i_1\ldots i_N)}.
\een

Now, in order to simplify the notation we assign:
\ben
i_1\ldots i_N\to \pmb i,\nonumber\\ k_1\ldots k_M \to \pmb k,\nonumber\\N!|\omega_{i_1\ldots i_N}|^2\to d_{\pmb i},\nonumber\\g_{k_1\ldots k_M}^{(i_1\ldots i_N)}\to g_{\pmb k\pmb i}.
\een
With this notation Eq. (\ref{A9}) becomes
\ben
G_{k_1\ldots k_M}=G_{\pmb k}=\sum_{\pmb i} d_{\pmb i}g_{\pmb k\pmb i},
\een
the normalization condition reads

\ben
\label{normal}
\sum_{\pmb i} d_{\pmb i}=1,
\een
and $g_{\pmb k\pmb i}$ satisfies
\ben
\label{sumg}
\sum_{\pmb k}g_{\pmb k\pmb i}^2&=&\binom{N}{M}^{-1}.
\een

Let us now consider the sum of the squares of the diagonal elements

\ben
\label{bigdemo}
\sum_{\pmb k}G_{\pmb k}^2&=&\sum_{\pmb k}\left(\sum_{\pmb i} d_{\pmb i}g_{\pmb k\pmb i}\right)^2\nonumber\\&=&\sum_{\pmb k}\left\{\sum_{\pmb i} d_{\pmb i}^2g_{\pmb k\pmb i}^2+2\left(\sum_{\pmb i<\pmb{i'}}d_{\pmb i}d_{\pmb {i'}}g_{\pmb k\pmb i}g_{\pmb k\pmb {i'}}\right)\right\}\nonumber\\&=&\sum_{\pmb k}\left\{ \left(\sum_{\pmb i} d_{\pmb i}\left(1-\sum_{\pmb {i'}\neq\pmb i} d_{\pmb {i'}}\right)g_{\pmb k\pmb i}^2\right)+2 \left(\sum_{\pmb i<\pmb{i'}}d_{\pmb i}d_{\pmb {i'}}g_{\pmb k\pmb i}g_{\pmb k\pmb {i'}}\right)\right\}\nonumber\\&=&\sum_{\pmb k}\left\{ \left(\sum_{\pmb i} d_{\pmb i}g_{\pmb k\pmb i}^2\right)-\left(\sum_{\pmb {i'}\neq\pmb i} d_{\pmb i}d_{\pmb {i'}}g_{\pmb k\pmb i}^2\right)+2 \left(\sum_{\pmb i<\pmb{i'}}d_{\pmb i}d_{\pmb {i'}}g_{\pmb k\pmb i}g_{\pmb k\pmb {i'}}\right)\right\}\nonumber\\&=&\sum_{\pmb k}\left\{ \left(\sum_{\pmb i} d_{\pmb i}g_{\pmb k\pmb i}^2\right)-\left(\sum_{\pmb i<\pmb{i'}}d_{\pmb i}d_{\pmb {i'}}(g_{\pmb k\pmb i}^2+g_{\pmb k\pmb {i'}}^2-2g_{\pmb k\pmb i}g_{\pmb k\pmb {i'}})\right)\right\}\nonumber\\&=&\sum_{\pmb i} d_{\pmb i}\left(\sum_{\pmb k}g_{\pmb k\pmb i}^2\right)-\left\{\sum_{\pmb i<\pmb{i'}}d_{\pmb i}d_{\pmb {i'}}\sum_{\pmb k}(g_{\pmb k\pmb i}-g_{\pmb k\pmb {i'}})^2\right\}
\een

Using the relations (\ref{normal}) and (\ref{sumg}), we finally get

\ben
\label{demofinal}
\sum_{\pmb k}G_{\pmb k}^2=\binom{N}{M}^{-1}-\left\{\sum_{\pmb i<\pmb{i'}}d_{\pmb i}d_{\pmb {i'}}\sum_{\pmb k}(g_{\pmb k\pmb i}-g_{\pmb k\pmb {i'}})^2\right\}\leq\binom{N}{M}^{-1}.
\een

Since we did not impose any restriction on the single-particle basis $\{|i\rangle\}$, Eq. (\ref{demofinal}) holds for any basis. In particular, it holds for the eigenbasis of $\rho_M$, in which $\sum_{\pmb k}G_{\pmb k}^2=\textrm{Tr}\rho_M^2$. We have thus established the following inequality

\ben
\label{TraceInequality}
\textrm{Tr}\rho_M^2\leq\binom{N}{M}^{-1}.
\een

The only way for the equality sign to hold in \eqref{TraceInequality} is to have one of the $d_{\pmb i}$ equal to $1$ and the rest equal to $0$, meaning that there is only one term in the original expansion for $\ket{\Psi}$, Eq.~\eqref{GeneralFermionSate}. This implies that $\ket{\Psi}$ has Slater rank one, and can thus be expressed as one single Slater determinant.

\end{widetext}

\end{appendix}

\bibliographystyle{apsrev}
\bibliography{Master_Bibtex}

\end{document}